# Efficiency enhancement in regenerative amplifier free-electron lasers using a tapered undulator

Henry P. Freund[1,2*], P.J.M. van der Slot[3] and Patrick G. O'Shea[1,4]

**Recent progress in short wavelength free-electron lasers extending from the extreme ultraviolet through the x-ray regime have opened new avenues for industrial and research applications. A high-gain/low-Q oscillator, i.e., a regenerative amplifier free-electron laser (RAFEL), is one possible concept. In this paper, we present the first analysis of efficiency enhancement in a RAFEL with a long, tapered undulator line. For the present analysis, we consider a high average power EUV RAFEL at 13.5 nm and show that its' performance exceeds that of single-pass, tapered self-amplified spontaneous emission. This points the way to high power tapered undulator x-ray RAFELs.**

THE development of coherent extreme ultraviolet (EUV) and x-ray sources based upon free-electron lasers (FELs) has largely relied upon single-pass self-amplified spontaneous emission (SASE)[1-7] or a high-gain harmonic generation cascade from a long(er) wavelength seed.[8,9] However, efforts are under way to develop oscillator-based FELs at these short wavelengths in order to improve stability of the output power and spectra. FEL oscillators can be grouped into two categories: low-gain/high-*Q* oscillators which we refer to as FELOs, and high-gain/low-*Q* oscillators which are often referred to as Regenerative Amplifiers (RAFELs) and which is the subject of this paper.

There is an extensive literature dealing with RAFELs[10-14] spanning the spectrum from the infrared[12] through hard x-rays.[13,14] In this paper, we present the first analysis of efficiency enhancement in a RAFEL with a long, tapered undulator line. This has important implications for both high average power industrial applications in the ultraviolet spectrum and for hard x-ray fourth generation light sources. For the present analysis, we consider a high average power 13,5 nm high average power EUV RAFEL. Consideration of an x-ray RAFEL using diamond Bragg mirrors will be presented in a future work.

In simulating a RAFEL, we use MINERVA[15-17] and OPC.[18,19] In this process MINERVA simulates the interaction in the undulator which then "hands-off" the optical mode to OPC which propagates the field around the resonator and back to the undulator entrance at which point the returning optical pulse is "handed back" to MINERVA. This is repeated for as many passes as desired. MINERVA/OPC has been validated by comparison with the 10-kW upgrade experiment (FELO) at Thomas Jefferson National Accelerator Facility[20] where MINERVA/OPC recorded an output power of 14.52 kW[15] which compares well with the 14.3 ± 0.72 kW measured in the experiment.

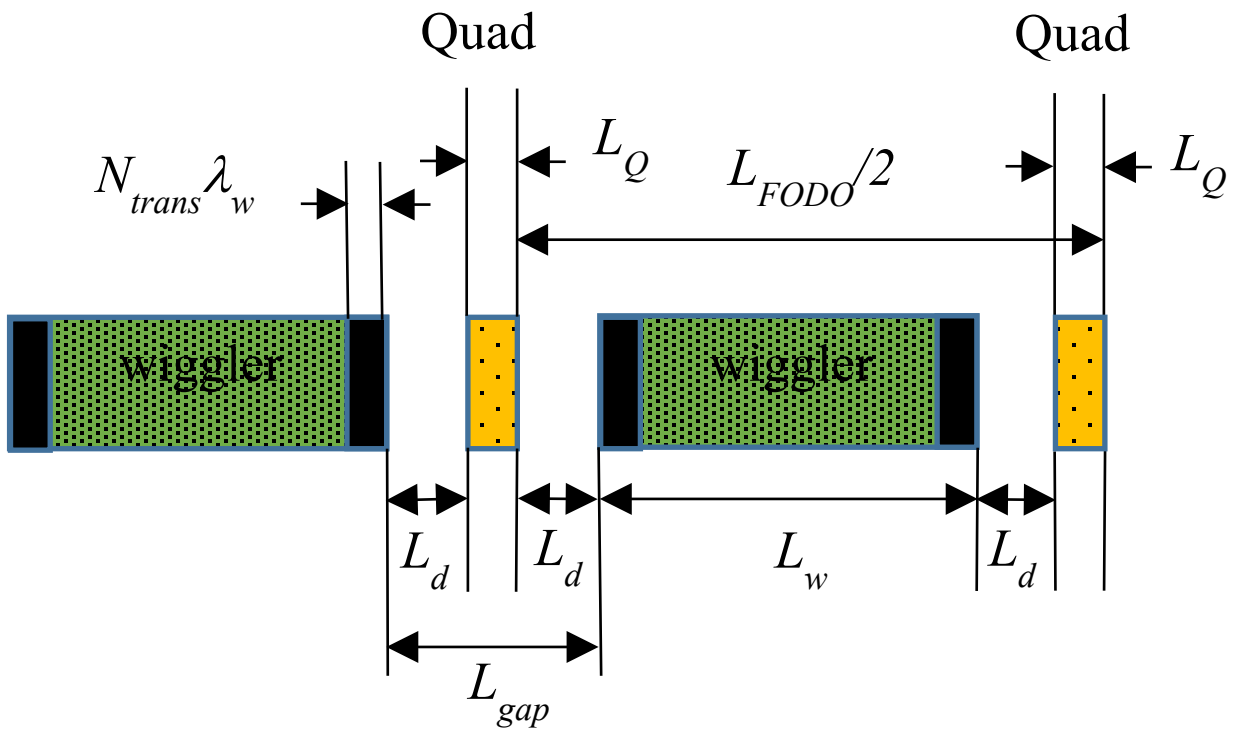

**Figure 1 | FODO cell layout.** Illustration of one FODO cell of the undulator line.

## The Physical Configuration

The basic parameters of the undulator/quadrupole lattice are listed in Table 1. The lattice we consider consists of an alternating sequence of undulators and quadrupoles comprising a focusing/defocusing (FODO) lattice as illustrated in Fig. 1 showing a single FODO cell consisting of two undulators and quadrupoles. The FODO cell length is 5.688 m. The undulators are plane polarized with flat-pole-faces with a peak on-axis field of 3.8142 kG a period of 3.5 cm and an overall length of 2.17 m (62 periods) including one period ($N_{trans}$ = 1) for the entry and exit transitions with 60 periods of uniform field. The quadrupoles are placed in the center of the gaps between the undulators and have a field gradient of 1.6 kG/cm and a length of 7.2 cm. The configurations under consideration in this paper will involve efficiency enhancements using a sequence of step-tapered undulators.

[1]Institute for Research in Electronics and Applied Physics, University of Maryland, College Park, MD 20742, USA; [2]NoVa Physical Science and Simulations, Vienna, Virginia 22182, USA; [3]Laser Physics and Nonlinear Optics, Mesa+Institute for Nanotechnology, University of Twente, Enschede, the Netherlands; [4]Department of Electrical Engineering, University of Maryland, College Park, Maryland 20742, USA; e-mail: hfreund@umd.edu

**Table 1 | Undulator, FODO cell parameters.**

| | |
|---|---|
| **Undulators: Planar, flat-pole-face** | |
| Period, $\lambda_w$ | 3.5 cm |
| Length, $L_w$ | 2.17 m |
| Entry/Exit transitions | 1 period |
| **Quadrupoles** | |
| Field Gradient | 1.6 kG/cm |
| Length, $L_Q$ | 7.2 cm |
| FODO Cell Length, $L_{FODO}$ | 5.66 m |

The electron beam parameters are listed in Table 2. The kinetic energy is 775 MeV, a peak current of 300 A with a parabolic temporal profile with a duration of 500 fs for a bunch charge of 100 pC, an emittance of 0.6 μm and an rms energy spread of 0.05%. Note that the rms bunch duration for a parabolic temporal profile with these parameters is 112 fs corresponding to an rms bunch length of about 33.6 μm. The bunch repetition rate is 50 MHz yielding an average current of 5 mA. The initial Twiss parameters are chosen to match the beam into the undulator/FODO lattice.

**Table 2 | Electron beam parameters.**

| | |
|---|---|
| Kinetic Energy | 775 MeV |
| Peak Current | 300 A |
| Pulse Duration (parabolic) | 500 fs |
| Bunch Charge | 100 pC |
| Repetition Rate | 50 MHz |
| Average Current | 5 mA |
| Normalized Emittance | 0.6 μm |
| rms Energy Spread | 0.05% |

**Table 3 | Resonator parameters.**

| | |
|---|---|
| Wavelength, $\lambda$ | 13.5 nm |
| Length (zero-detuning) | 110.02320946 m |
| Mirror Coating | Mo/Si multilayer |
| Mirror Reflectivity | 70% |
| Mirror Curvature | 75.0 m |
| Mirror Radius | 3.5 cm |
| Hole Radius | 0.7 – 1.2 mm |

These undulator/beam parameters yield an FEL resonance at a wavelength of 13.5 nm. The parameters of the resonator for the RAFEL is chosen with this in mind. We use a stable, concentric resonator with a (tunable) length in the vicinity of 110.02320946 m corresponding to the zero-detuning length ($L_0 = c/2f_{rep}$). The basic parameters are shown in Table 3. The radii of curvature of the mirrors is 75.0 m with a mirror radius of 3.5 cm. We assume Mo/Si multilayer mirror coatings for which the mirror reflectivity is about 70% at this wavelength.[21,22]

Out-coupling is through a hole in the downstream mirror and this will ensure that mirror loading is kept within acceptable limits the damage threshold. We will discuss this in more detail below.

In studying the FEL performance envelope for this configuration, we are interested in limits on the output power and the optical stability with respect to the power and spectrum (wavelength and relative bandwidth). On the practical side, the assumption of a 50 MHz repetition rate implies that the accelerator will be a superconducting energy recovery linear accelerator (ERL)[23] so that the energy distribution of the spent beam is an important consideration in the stable operation of the ERL. However, while fluctuations in the injected electron beam (energy, emittance, energy spread, slew, etc.) are important considerations in the design of an actual system, they are beyond the scope of this paper.

## Self-Amplified Spontaneous Emission (SASE)

As a point of comparison for the RAFEL, we first consider the performance of an optimized, step-tapered undulator line operating in single-pass Self-Amplified Spontaneous Emission (SASE) mode. In SASE simulations, we employ an ensemble of 20 different noise seeds to generate the initial phase space, and we optimize the taper in both the first tapered undulator and the (linear) slope of the subsequent down taper. The optimal taper configuration is found to start at the 14th undulator and decrease thereafter with a slope of $\Delta B_w/B_w = -0.26\%$ for an additional 12 tapered undulators and an overall length of 70 m.

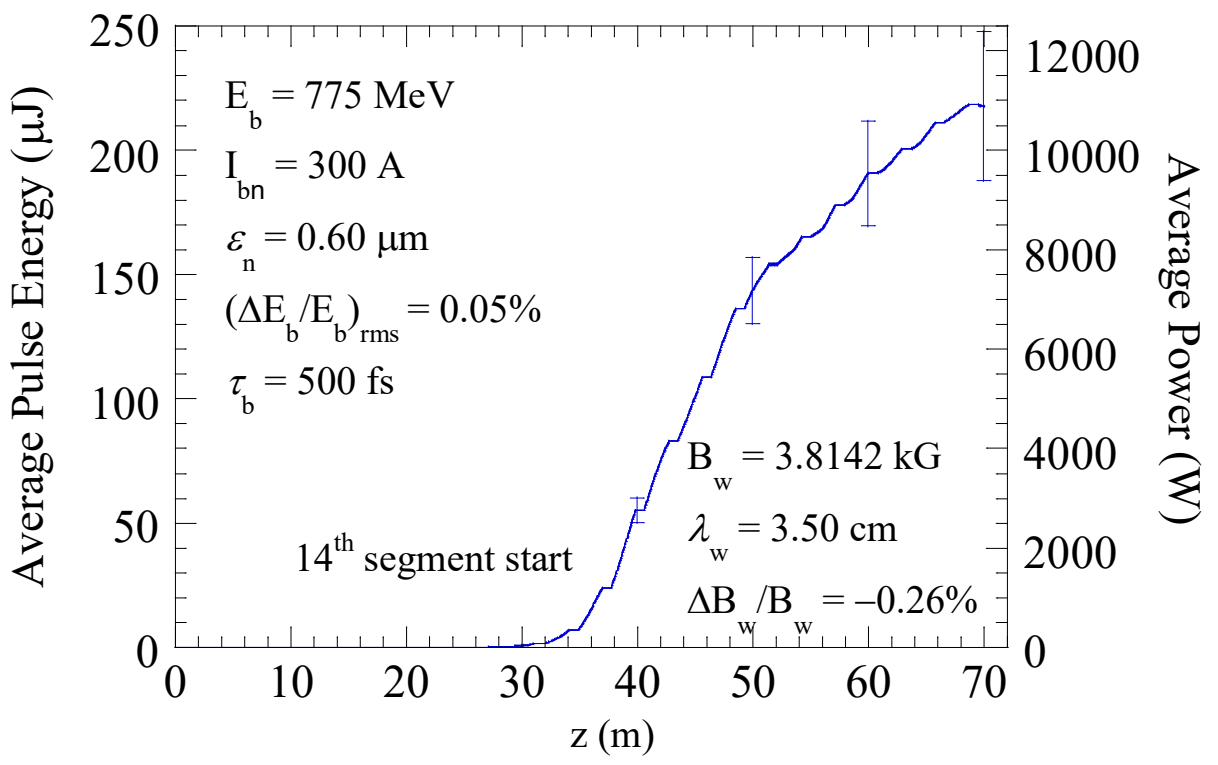

**Figure 2 | SASE Performance.** Evolution of the pulse energy (left) and corresponding average Power (right) for the optimized taper.

Time-dependent simulations of the optimized taper configuration are shown in Fig. 2 where we plot the pulse energy (left axis) and the corresponding average power (right axis). As shown in the figure, the average peak pulse energy after 70 m reaches a level of approximately 218 μJ corresponding to an average power of about 10.8 kW. This represents an increase over simulations of a uniform undulator line by a factor of about 2.7. This is consistent with what is found in tapered undulator amplifiers.[24] It is evident, that the pulse energy is beginning to roll over as the length of the tapered undulator increases to 70 m and we expect that the performance will be degraded by slippage for still longer undulator lines. This is a limiting factor in the performance of all long undulator/single-pass tapered undulator FELs. The shot-to-shot fluctuation level at the end of the undulator line is 29.9 μJ (14%) and this is consistent with expectations for SASE.

The characteristics of the SASE output spectrum are of particular importance. The average wavelength (blue, left axis) and relative linewidth (red, right axis) are shown in Fig. 3. As shown, the average wavelength (13.506 nm) is relatively stable and exhibits an rms fluctuation level of 0.03%. The rms value of the relative linewidth is approximately 0.15% with a fluctuation of about 22% from shot-to-shot.

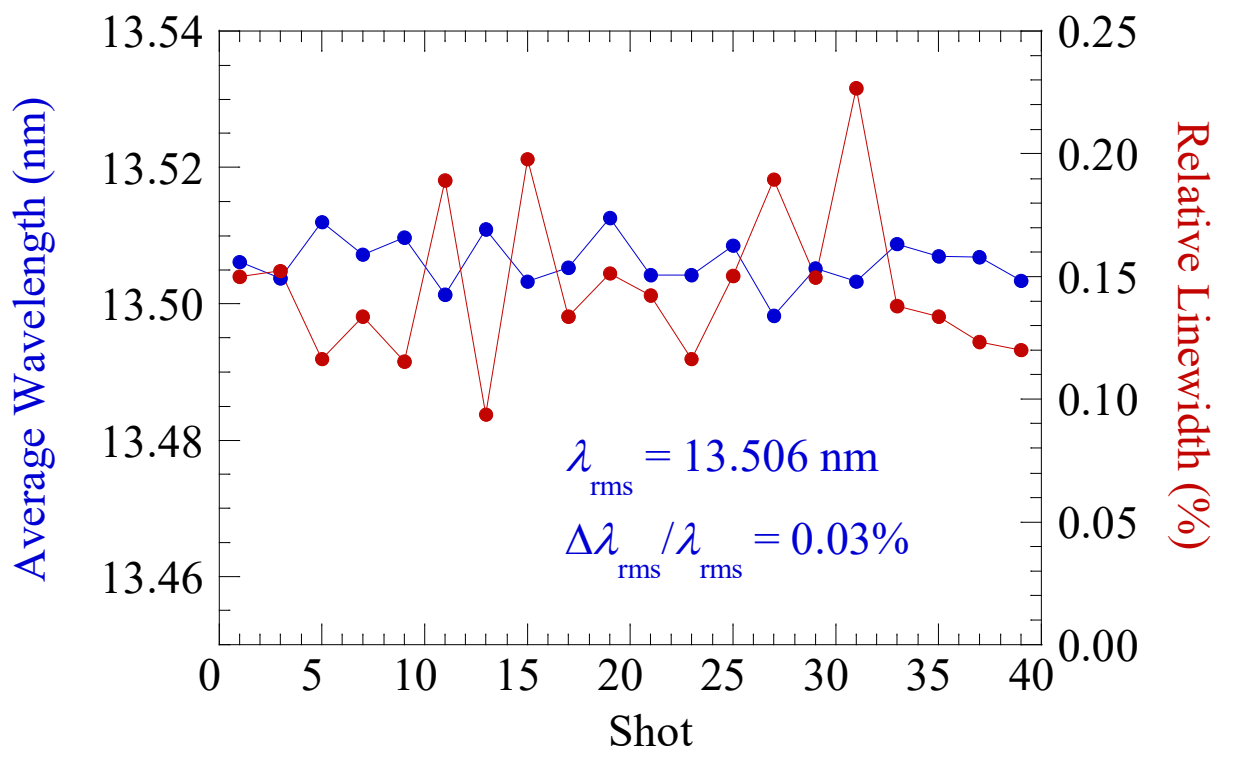

**Figure 3 | SASE Stability.** Shot-to-shot evolution of the average wavelength and the relative linewidth.

The spent beam distribution is shown in Fig.4. As indicated in the figure, the rms energy spread at the undulator exit is 0.475% which is about an order of magnitude greater than that of the beam upon entry to the undulator. As pointed out by Hajima,[25] the energy acceptance of the spent beam is a crucial factor in the stability of the ERL and the design of the returning arcs are an important component to the ERL design. The ERL used in the 10-kW upgrade experiment at the Thomas Jefferson National Accelerator Facility[20] employed a Bates bend[26] which had an energy acceptance of 15%. Hence, we do not expect that the spent beam energy spread found in simulation would destabilize the ERL.

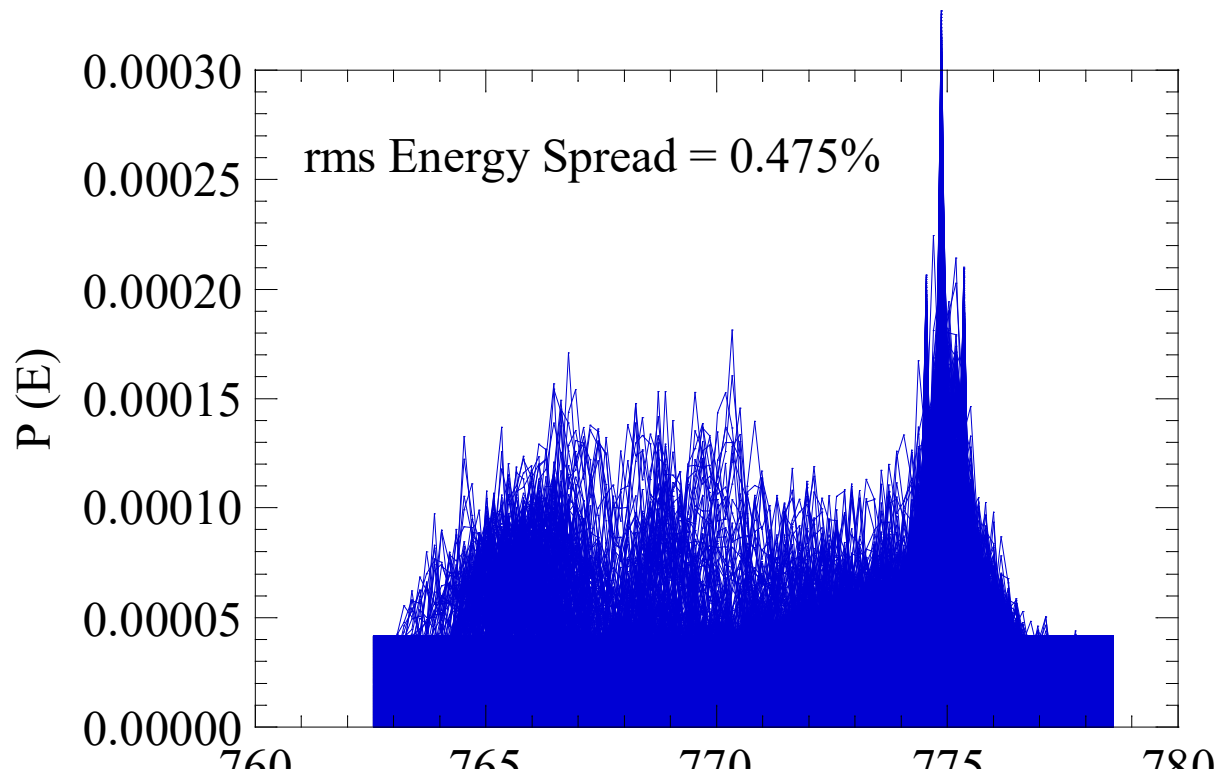

**Figure 4 | SASE Spent beam.** The spent beam distribution at the undulator exit.

## Regenerative amplifier (RAFEL) performance

Optimization of a tapered-undulator RAFEL requires simultaneous, multi-dimensional parameter scans. Since it is important to minimize thermal loading on the mirrors, we employ hole out-coupling on the downstream mirror. Since the bulk of the energy incident on this mirror will pass through the hole, and since there are substantial mirror losses, both the hole radius and the gain in the undulator must be simultaneously optimized. In particular, the undulator line begins with a uniform section which is followed by the step-down-taper and the number of undulators in each segment, as well as the position of the first tapered segment and the slope of the taper, must be optimized to obtain an adequate gain in the saturated state to overcome the losses for each choice of hole radius. Of course, this also involves a scan over cavity lengths to obtain the detuning curve.

For the configuration under consideration, the optimal hole radius is found to be 1.1 mm which out-couples about 95% of the incident power. An additional roundtrip loss of 4% is due to mirror absorption. The undulator configuration must be designed so that the single-pass gain in the saturated regime replenishes these losses. This is accomplished using an undulator line with an overall length of 57.2 m consisting of 8 undulators in the uniform section and 12 undulators in the tapered section with a slope of $\Delta B_w/B_w = -0.30\%$, which is slightly steeper than that found for the SASE case discussed above. We find that the average gain per pass in the saturated regime is approximately $10^5$%.

The detuning curve is shown in Fig. 5 where we plot the output optical pulse energy (left axis) and the average output power (right axis) versus the normalized detuning length relative to the zero-detuning length, $L_0$ (= $c/2f_{rep}$).

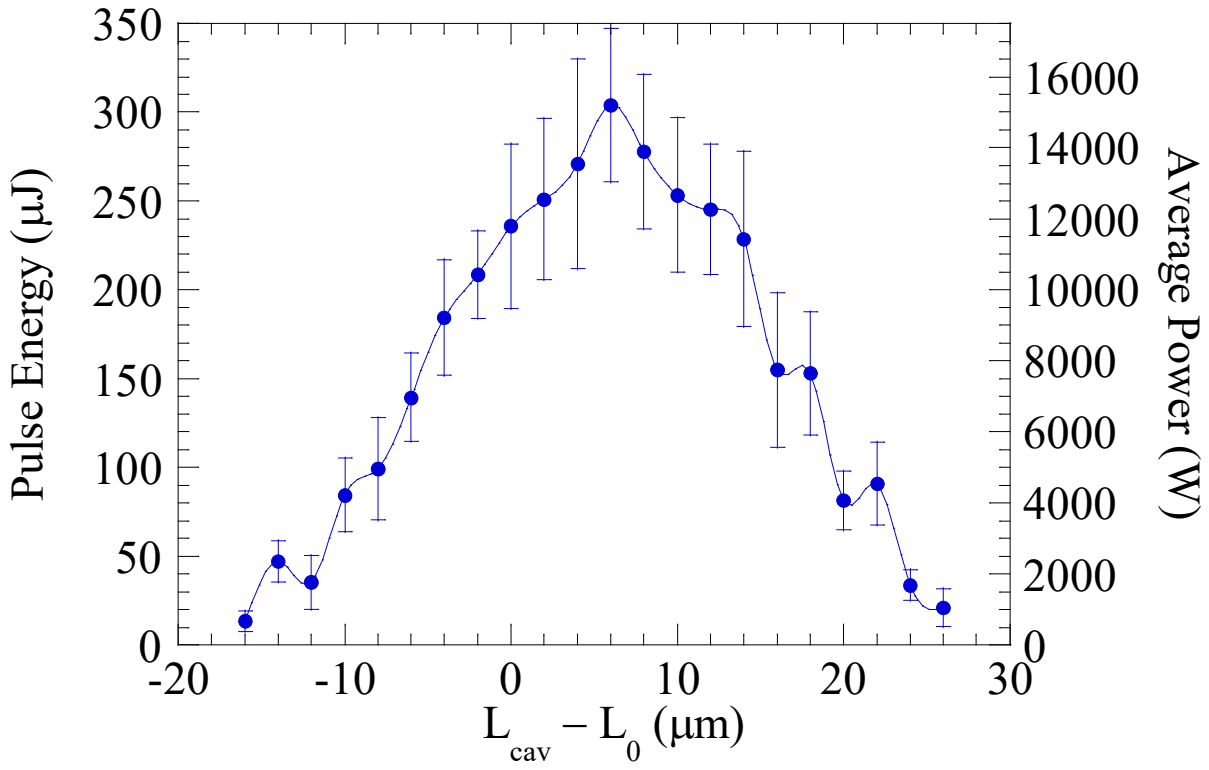

**Figure 5 | RAFEL Detuning curve.** The detuning curve showing the level of pass-to-pass fluctuations.

There are several points to note from the detuning curve:

1. At the peak of the detuning curve, the output pulse energy is about 304 μJ which corresponds to an average power of 15.2 kW. This is higher than found for single-pass SASE.
2. The peak of the detuning curve is found for a positive detuning of 6 μm. This indicates that the optimum cavity length is longer than the zero-detuning length and the returning optical pulse slightly lags the incoming electron bunch. As a result, the optical pulse will slip along (but not off) of the electron bunch over the course of the undulator line but will primarily interact with the peak of the electron bunch. This ameliorates the effect of slippage which is the limiting factor in single-pass SASE. This is also in marked contrast to FELOs where the peak of the detuning curve is found to be close to the zero-detuning length.
3. The width of the detuning curve is about 42 μm. The rms bunch length is approximately 34 μm so that the detuning curve represents the pass-to-pass response governed largely by the peak of the electron bunch. This width is much wider than what is expected in an FELO because the gain per pass in the saturated regime is extremely high. This contrasts with the gain per pass in the saturated regime in the 10-kW Upgrade experiment at Jefferson Laboratory of about 22% which compensated for the out-coupling and mirror losses.[27]
4. The pass-to-pass fluctuation level at the peak of the detuning curve is approximately 14% which is comparable to the fluctuation level found for shot-to-shot single-pass SASE. The fluctuations in the RAFEL are due to limit cycle oscillations which are found in both FELOs[28] and in RAFELs.[15]

Not shown in the figure is that there is a "floor" to the detuning curve as the cavity length either increases or decreases to the point where there is no overlap between the optical pulses and the electron bunches because the single pass SASE emission is of the order of 1 μJ.

The evolution of the pulse energy and average power versus pass is shown in Fig. 6 for the cavity length of $L_{cav}$ = 110.92321546 m corresponding to the peak in the detuning curve. As indicated in the figure, the saturation regime is reached after about 4 passes at a pulse energy/average power of 304 μJ/15.2 kW. The rms pass-to-pass fluctuation level is about 14.3%. This is comparable what we found in the steady-state simulation.

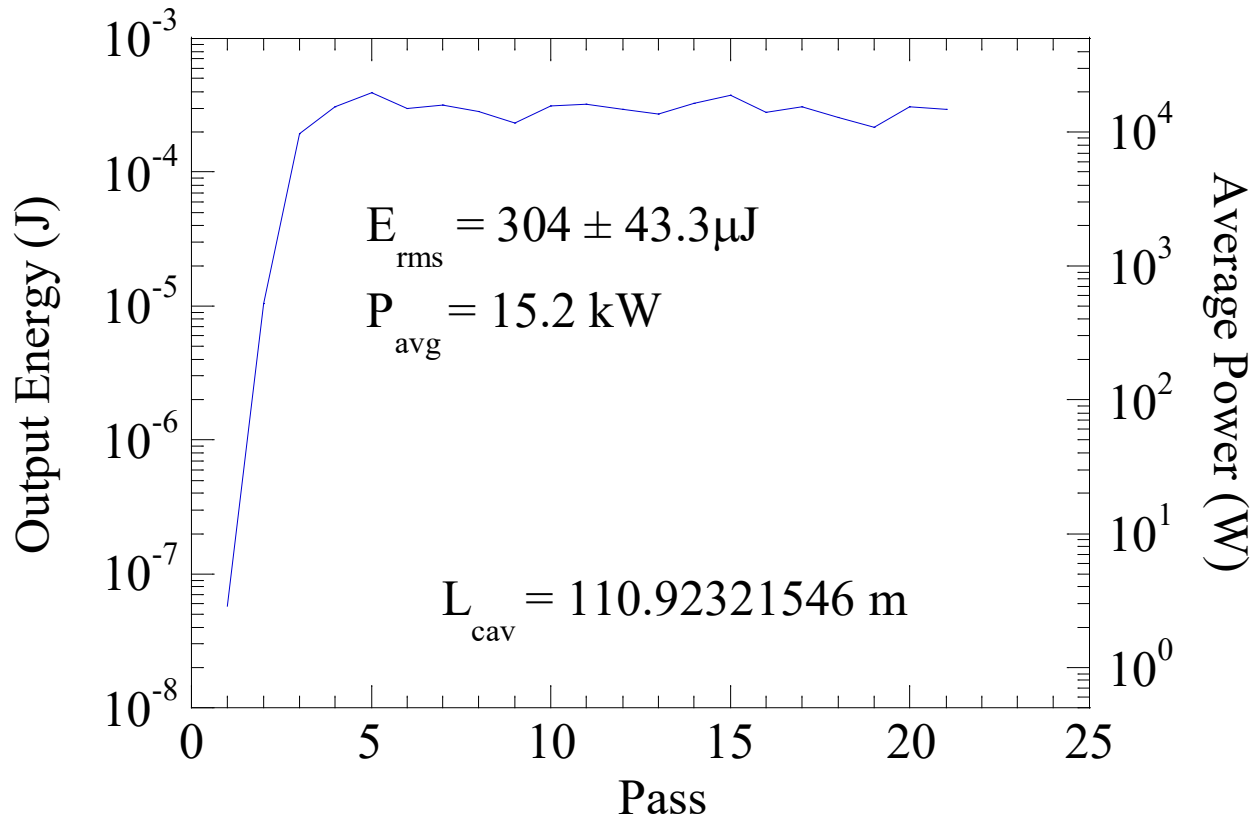

**Figure 6 | RAFEL Peak performance.** Pass-to-pass evolution of the RAFEL performance at the peak in the detuning curve.

About 95% of the incident power on the downstream mirror is outcoupled through the hole. Given a reflection coefficient of 70%, the average power absorbed on the downstream mirror is 169 W while that absorbed on the upstream mirror is 119 W. The damage threshold at 13.5 nm for the peak power density on Mo/Si multilayer mirrors has been reported[21,22] to be 83 mJ/cm$^2$. The rms width of the incident pulse on the downstream mirror is found to be 0.044 cm, so that using the rms pulse duration (112 fs) the peak energy density incident on the downstream mirror is about 37 mJ/cm$^2$ which over estimates the peak energy density since the majority of the pulse that is incident on the mirror is outcoupled through the hole. The peak power incident on the upstream mirror is about 70 .7 MW and the rms width of the pulse is about 0.366 cm so that the peak energy density is about 19 μJ/cm$^2$. These are well below the reported damage threshold and should be manageable.

The characteristics of the output spectrum in the RAFEL are of particular importance. The average wavelength (blue, left axis) and relative linewidth (red, right axis) are shown in Fig. 7 at the optimal detuning. As shown the average wavelength is relatively stable in the saturated regime (13.512 nm) and exhibits an rms fluctuation level of 0.05%. The rms value of the relative linewidth is approximately 0.522 with a fluctuation of about 9% from pass-to-pass. Note that the fluctuation level and the relative linewidth are somewhat larger than the shot-to-shot values for the SASE simulations.

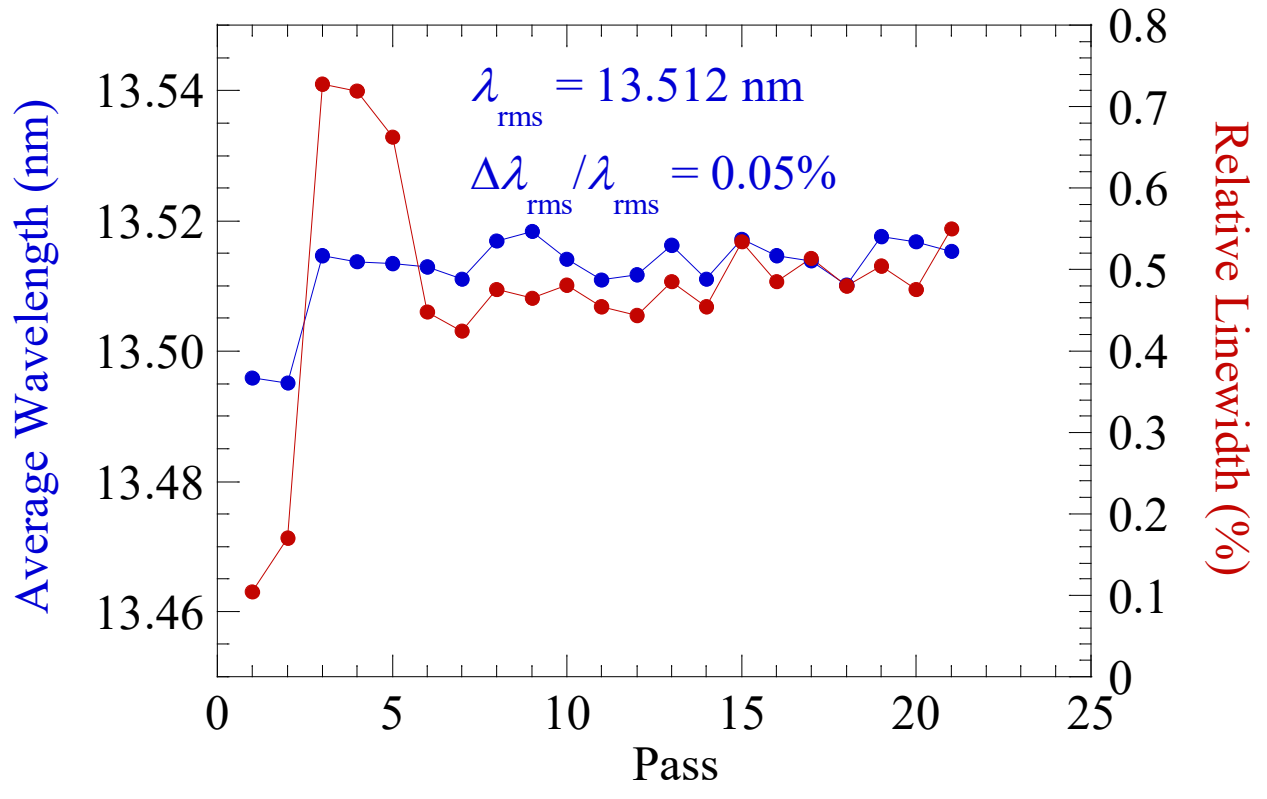

**Figure 7 | RAFEL Spectral Stability.** Pass-to-pass evolution of the average wavelength and relative linewidth at the peak in the detuning curve.

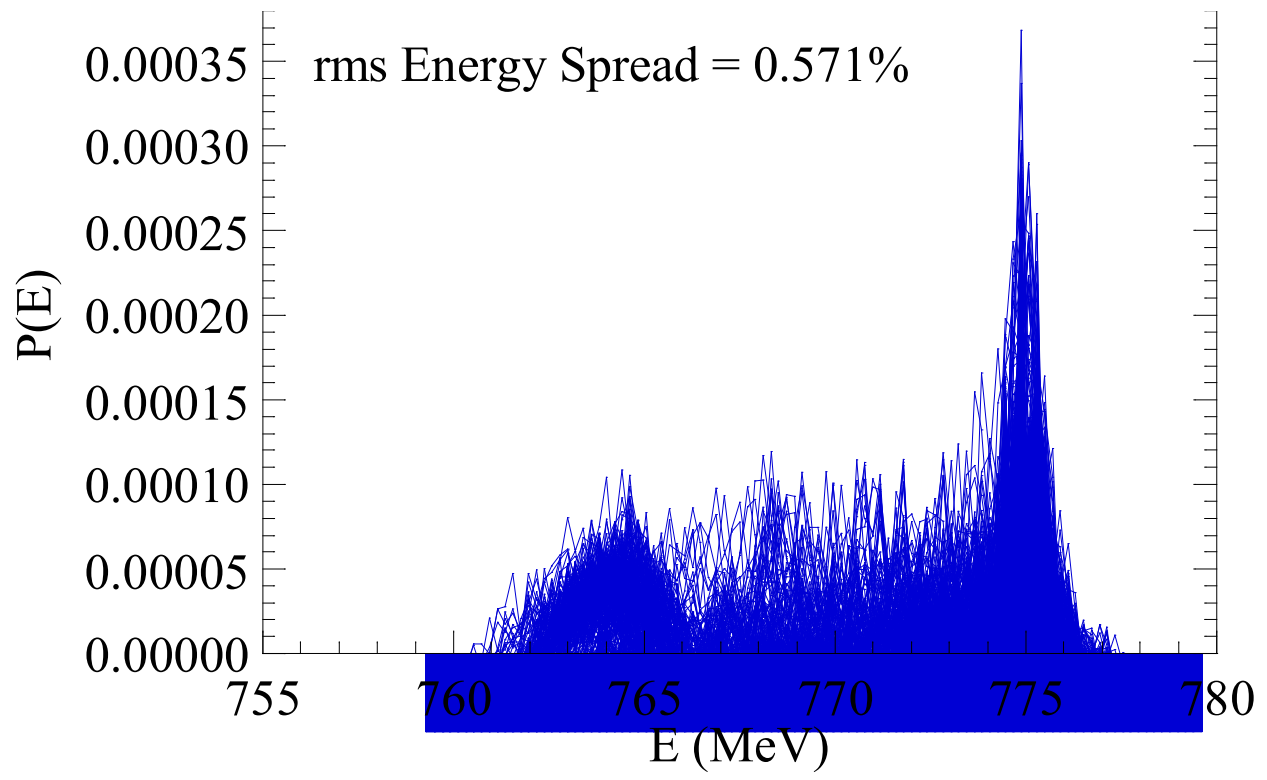

**Figure 8 | RAFEL Spent beam.** The spent beam energy distribution at the peak in the detuning curve.

The spent beam distribution from a pass in the saturated regime is shown in Fig. 8. As shown, the rms energy spread is 0.571% which is somewhat larger than found for the single-pass SASE case. This is consistent with the somewhat higher output power for the RAFEL. The spent beam energy spread found in the RAFEL is comparable to that in the single-pass SASE and is not expected to pose a problem for the stability of the ERL.

## Summary and Discussion

In this paper, we present the first analysis of efficiency enhancement in a RAFEL with a long, tapered undulator line, and apply the nonlinear simulation to an example of a high average power 13.5 nm RAFEL. In comparison with single-pass SASE in a tapered undulator where the limiting factor is slippage over the long undulator, recirculation in the RAFEL allows for a resonator in which the returning optical pulse lags the incoming electron bunch so that the optical pulse slips over the peak in the electron bunch and not off the peak. This minimizes the deleterious effect of slippage. As such, the performance of the RAFEL exceeds that of single-pass SASE. However, the pass-to-pass fluctuations in the RAFEL are comparable to the shot-to-shot fluctuations in SASE and the wavelength and spectral linewidth fluctuations in the RAFEL are also comparable to what is found in SASE.

The present study points the way to high-power x-ray RAFELs which may be implemented in $4^{th}$ generation light source facilities such as the LCLS-II at the Stanford Linear Accelerator Laboratory[29] which, instead of an ERL, is based upon a high-energy superconducting linac with a high repetition rate. Such RAFELs are under consideration at the present time.[30] Simulations of such an x-ray RAFEL with a tapered undulator line will be presented in a forthcoming publication.

## Acknowledgements

This work was supported by the United States Department of Energy under contract DE-SC0024397. The authors would like to thank Dr. S.V. Benson for many helpful discussions.